\documentclass{ws-mpla}
\usepackage[super]{cite}
\usepackage{graphicx}

\usepackage{dcolumn}
\usepackage{bm}
\usepackage{amsmath}
\usepackage{epstopdf}
\usepackage{amsfonts}
\usepackage{amssymb}
\usepackage{epsfig}
\usepackage{tabularx}
\usepackage{color}
\usepackage{hyperref}

\begin{document}

\markboth{Grigoris Panotopoulos}
{Electromagnetic quasinormal modes of the nearly-extremal higher-dimensional Schwarzschild-de Sitter black hole}

\catchline{}{}{}{}{}

\title{Electromagnetic quasinormal modes of the nearly-extremal higher-dimensional Schwarzschild-de Sitter black hole}

\author{\footnotesize Grigoris Panotopoulos 
}

\address{
Centro de Astrof{\'i}sica e Gravita{\c c}{\~a}o-CENTRA, Departamento de F{\'i}sica, 
\\
Instituto Superior T\'ecnico-IST,
Universidade de Lisboa-UL,
\\
Av. Rovisco Pais 1, 1049-001 Lisboa, Portugal
%
\\
\href{mailto:grigorios.panotopoulos@tecnico.ulisboa.pt}{\nolinkurl{grigorios.panotopoulos@tecnico.ulisboa.pt}}
}

\maketitle

\pub{Received (Day Month Year)}{Revised (Day Month Year)}

\begin{abstract}
We obtain an analytical expression for the electromagnetic quasinormal spectrum of the higher-dimensional nearly-extremal Schwarzschild-de Sitter black hole. The WKB method is used to verify the results, and a comparison with known results from previous works is briefly made as well.

\end{abstract}

\ccode{PACS Number(s): 04.30.Nk, 04.50.Gh, 04.70.Bw}

\section{Introduction}

Quasinormal (QN) modes characterize the dumbed oscillations of the geometry of space time when a black hole is perturbed, and they give us information about the stability of these perturbations \cite{wheeler,zerilli1,zerilli2,zerilli3,moncrief,teukolsky,monograph}. Recently they attracted a lot of renewed interest thanks to the direct detection of gravity waves seen by the LIGO collaboration \cite{ligo1,ligo2,ligo3}. Gravity wave Astronomy may test gravity and probe strong gravitational fields, while by observing the quasinormal spectra, at least in principle, we may determine the black hole parameters, such as mass, charge and angular momentum. However, one should bear in mind that as it has been pointed out e.g. in \cite{Yunes,Roman}, the uncertainty in mass and angular momentum is large, leaving space for alternative theories. For reviews on the topic see e.g. \cite{review1,review2,review3}.

The current cosmic acceleration \cite{sn1,sn2} as well as the AdS/CFT correspondence \cite{adscft1,adscft2} motivate the study of spacetimes with a non-vanishing cosmological constant. Furthermore, Kaluza-Klein theory \cite{kaluza,klein}, supergravity \cite{nilles} and Superstring Theory \cite{ST1,ST2} suggest and/or require that extra spatial dimensions may exist. It thus becomes clear that it would be interesting to see what kind of quasinormal spectra one expects to see when a higher-dimensional de Sitter or anti-de Sitter black hole is perturbed.

The topic of QN modes is undoubtedly a very active and rich field. Here, however, we shall not consider cases such as purely Schwarzschild BHs, purely de Sitter BHs, rotating BHs or alternative theories of gravity. The quasinormal frequencies of the Schwarzschild-de Sitter in 4 dimensions have been studied e.g. in \cite{basic,paper2,paper3} and references therein (see also e.g. \cite{paper4,paper5} and references therein for works on Schwarzschild-anti-de Sitter). 
In particular, in \cite{basic} an analytical formula for the QN modes was obtained for the nearly-extremal black hole, in \cite{paper2} the QN frequencies were computed employing the WKB semi-analytical approximation \cite{wkb1,wkb2}, and in \cite{paper3} the authors computed the high overtones of the 4D Schwarzschild-de Sitter black hole. 

In higher dimensions, $D > 4$, QN modes have been studied e.g. in \cite{KZ,molina,master,ortega1,ortega2,ortega3} and references therein. An extensive search of QNMs of the higher-dimensional Reissner-Nordstr{\"o}m-de Sitter black hole, both in time and in frequency domain, was realized in \cite{KZ}. An analytical formula for the QN spectrum for scalar perturbations of the higher-dimensional Schwarzschild-de Sitter black hole in the near extremal case, similar to the one found in \cite{basic}, has been obtained in \cite{molina}. In \cite{master} the extension to higher dimensions of the Zerilli equation for polar (even parity) and of the Regge-Wheeler equation for axial (odd parity) gravitational perturbations  was given, and the QN spectrum for gravitational perturbations of the nearly-extremal higher-dimensional Schwarzschild-de Sitter black hole can be found in \cite{review2}. However, to the best of our knowledge, an analytical expression for the QN spectrum of the nearly-extremal higher-dimensional Schwarzschild-de Sitter black hole for electromagnetic perturbations is still missing.

It is precisely the goal of this short article to fill this gap in the literature. Therefore, here we compute for the first time the QN modes of the nearly-extremal Schwarzschild-de Sitter black hole in $D > 4$ dimensions for electromagnetic perturbations. Our work is organized as follows: In the next section we present the structure horizon of the black hole as well as the effective potential that enter into the Scr{\"o}dinger-like equation, and in section 3 we obtain an analytical expression for the quasinormal spectrum in the near extremal case. We compare with results obtained within the WKB semi-classical approximation as well. We summarize our work in the fourth section. We set Newton's constant to unity, $G_N=1$, and we adopt the mostly positive metric signature $(-,+,...,+)$.

\section{The higher-dimensional black hole}

We consider Einstein's field equations in vacuum with a positive cosmological constant $\Lambda_D$ in a $D$-dimensional space time
\begin{equation}
R_{\mu \nu} - (1/2) R g_{\mu \nu} + \Lambda_D g_{\mu \nu} = 0
\end{equation}
with $g_{\mu \nu}$ being the metric tensor, $R_{\mu \nu}$ the Ricci tensor and $R$ the Ricci scalar.
The maximally symmetric solution of the field equations is the Schwarzschild-de Sitter solution given by
\begin{equation}
ds^2 = - f(r) dt^2 + f(r)^{-1} dr^2 + r^2 d \Omega_{D-2}^2
\end{equation}
where $r$ is the radial coordinate, and $\Omega_{D-2}^2$ is the line element of the unit $D-2$-sphere, while the metric function is given by
\begin{equation}
f(r) = 1-\frac{2 M}{r^{D-3}} - \frac{r^2}{a^2}=-\frac{r^{D-1}-a^2 r^{D-3}+2Ma^2}{a^2 r^{D-3}}
\end{equation}
where $M$ is the mass of the black hole, and $a$ is a length scale related to the cosmological constant, $\Lambda_D = (D-1) (D-2)/(2 a^2)$. To study the horizon structure one has to find the roots of the equation $f(r)=0$. Provided that
\begin{equation}
\frac{M^2}{a^{2(D-3)}} < \frac{(D-3)^{D-3}}{(D-1)^{D-1}}
\end{equation}
there are i) two real positive roots corresponding to an event horizon $r_H$, and a cosmological horizon, $r_c > r_H$, and ii) a bunch of more roots, either negative or complex, $r_i$, with the index $i=1,2,..,D-3$. Therefore the metric function can be factorized as follows
\begin{equation}
f(r) =- \frac{(r-r_H) (r-r_c) (r-r1) ... (r-r_{D-3})}{a^2 r^{D-3}}
\end{equation}

To compute the QN frequencies $\omega_n$ of electromagnetic perturbations, one has to solve a Scr{\"o}dinger-like equation of the form
\begin{equation}
\frac{d^2 \psi}{d r_*^2} + (\omega^2 - V_{S,V}) \psi(r_*) = 0
\end{equation}
where the so-called tortoise coordinate $r_*$ is given by
\begin{equation}
r_* = \int \frac{dr}{f(r)}
\end{equation}
The effective potentials $V_{S,V}(r_*)$ for electromagnetic perturbations are of two types, namely scalar-type (S) and vector-type (V), and they are given by \cite{ortega1,crispino}
\begin{equation}
V_{S}(r)  =  f(r) \left[ \frac{l (l+D-3)}{r^2} + \frac{(D-2) (D-4) f(r)}{4 r^2} - \frac{(D-4)f'(r)}{2r} \right]
\end{equation}
and
\begin{equation}
V_{V}(r)  =  f(r) \left[ \frac{(l+1) (l+D-4)}{r^2} + \frac{(D-6) (D-4) f(r)}{4 r^2} + \frac{(D-4)f'(r)}{2r} \right]
\end{equation}
where the prime denotes differentiation with respect to $r$. When $D=4$ both expressions collapse to the well-known case $V_{EM}=f(r) [l (l+1)/r^2]$, where $l$ is the degree of angular momentum.

\section{The spectrum in the near extremal case}

In this section we show that when $r_H \simeq r_c$ the effective potential that enters into the Scr{\"o}dinger-like equation takes the P{\"o}schl-Teller form \cite{potential}
\begin{equation}
V(x) = \frac{V_0}{cosh^2(\alpha(x-x_0))}
\end{equation}
with $V_0$ being the height of the potential and $\alpha$ being its width. It is known that for a potential of this form one can obtain an exact analytical expression for the quasinormal frequencies \cite{ferrari}
\begin{equation}
\omega_n = \pm \sqrt{V_0 - (\alpha/2)^2} - \alpha (n+1/2) i
\end{equation}
where $n=0,1,2,...$ is the overtone number. We can rewrite the previous formula equivalently as follows
\begin{equation}
\frac{\omega_n}{\alpha} = \pm \sqrt{V_0/\alpha^2 - 1/4} - (n+1/2) i
\end{equation}

To prove this we closely follow the notation of \cite{basic,molina}. First, since $r_H < r < r_c$, when $r_c \simeq r_H$ 
the radial coordinate is approximately constant $r \simeq r_H$, and the factorized form of the metric function $f(r)$ becomes approximately
\begin{equation}
f(r) \simeq A (r-r_H) (r_c-r)
\end{equation}
where the constant $A$ is related to the surface gravity $\kappa = (1/2) f'(r_H)$ through the simple relation $A (r_c-r_H) = 2 \kappa$.
Then the integral for the tortoise coordinate can be performed exactly finding $r$ as a function of $r_*$ in a closed form
\begin{equation}
r = \frac{r_H + r_c e^{2 \kappa r_*}}{1 + e^{2 \kappa r_*}}
\end{equation}
Therefore, $f(r)$ as a function of the tortoise coordinate is found to be
\begin{equation}
f(r_*) = \frac{(r_c-r_H) \kappa}{2 cosh^2(\kappa r_*)} 
\end{equation}
Finally, in the expressions for the effective potentials we observe that when $r_c \simeq r_H$, both the metric function and its derivative are negligible compared to the angular momentum term, which is the dominant one.
Therefore, the effective potential for the perturbations takes approximately the form of the P{\"o}schl-Teller form with width $\alpha=\kappa$ and height
\begin{equation}
V_0^{V} = \frac{\kappa (r_c-r_H)(l+1) (l+D-4)}{2 r_H^2}
\end{equation}
for vector-type, and
\begin{equation}
V_0^{S} = \frac{\kappa (r_c-r_H) l (l+D-3)}{2 r_H^2}
\end{equation}
for scalar-type.

The main result of this short article is thus summarized in the following two formulas
\begin{equation}
\boxed{\frac{\omega_{n,l}^{V}}{\kappa} = \sqrt{\frac{(r_c-r_H) (l+1) (l+D-4)}{2 \kappa r_H^2} - 1/4} - (n+1/2) i}
\end{equation}
for vector-type, and
\begin{equation}
\boxed{\frac{\omega_{n,l}^S}{\kappa} = \sqrt{\frac{(r_c-r_H) l (l+D-3)}{2 \kappa r_H^2} - 1/4} - (n+1/2) i}
\end{equation}
for scalar-type, while the surface gravity is terms of the parameters $M, a$ is computed to be \cite{molina}
\begin{equation}
\kappa = \left( \frac{1}{a^2} - \frac{M^2}{a^{2(D-2)}} \frac{(D-1)^{D-1}}{(D-3)^{D-3}} \right)^{1/2}
\end{equation}

We see that the real part of the modes depends on the degree of angular momentum only, while the imaginary part of the frequencies depends entirely on the overtone number. Furthermore, for $D=4$ we recover the spectrum for scalar and electromagnetic perturbations studied in \cite{basic}, while the spectrum for scalar-type and for arbitrary $D$ coincides with the spectrum for scalar perturbations studied in \cite{molina}.


As an example we consider the case where $D=5, M=1, a=2.829$. The horizons and the surface gravity are computed to be $r_H=1.98, r_c=2.02$ and $\kappa=0.01$. The real and imaginary part of the modes for vector-type are shown in Table~\ref{table:Spectrum} for $l=1$ up to $l=10$ and for $n=0$ up to $n=9$. 

\begin{table}[ht!]
\tbl{Real and imaginary part of the QNMs (vector-type) for $D=5, M=1, a=2.829$.}
{
\begin{tabular}{l | l | l | l}
		Degree $l$ & $Re(\omega)$ & Overtone $n$ & $Im(\omega)$ \\
		\hline
		\hline
		1 & 0.0135 & 0 & -0.0051  \\
		2 & 0.0211 & 1 & -0.0152  \\
		3 & 0.0284 & 2 & -0.0254  \\
		4 & 0.0357 & 3 & -0.0356  \\
		5 & 0.0430 & 4 & -0.0457  \\
		6 & 0.0503 & 5 & -0.0559  \\
		7 & 0.0575 & 6 & -0.0660  \\
		8 & 0.0648 & 7 & -0.0762  \\
		9 & 0.0720 & 8 & -0.0864  \\
		10 & 0.0793 & 9 & -0.0965  \\
\end{tabular}
	\label{table:Spectrum}
}
\end{table}

\subsection{Comparison with the WKB method}

To verify our results we make comparison with the semi-analytical WKB approximation \cite{wkb1,wkb2}, which in the eikonal limit ($l \rightarrow \infty$) becomes increasingly accurate. Following the ideas and the formalism developed in \cite{eikonal}, in the eikonal limit the QN modes of black holes in any spacetime dimensionality are determined by the parameters of the circular null geodesics
\begin{equation} \label{eikonal}
\omega_{l \gg 1} = \Omega_c l - i \left(n+\frac{1}{2}\right) |\lambda_L|
\end{equation}
where the Lyapunov exponent $\lambda_L$ is given by \cite{eikonal}
\begin{equation}
\lambda_L = \sqrt{\frac{1}{2}f(r_1) r_1^2 \left( \frac{d^2}{dr^2} \frac{f}{r^2}  \right)\bigg{|}_{r=r_1}}
\end{equation}
while the angular velocity $\Omega_c$ at the unstable null geodesic is given by \cite{eikonal}
\begin{equation}
\Omega_c = \sqrt{\frac{f(r_1)}{r_1^2}}
\end{equation}
with $f(r)$ being the metric function, and $r_1$ is the root of the algebraic equation \cite{eikonal}
\begin{equation} \label{algebraic}
2 f(r_1) - r_1 f'(r)|_{r_1} = 0
\end{equation}
It is straightforward to verify that precisely the same expression for the QN frequencies in the eikonal regime can be obtained by the WKB method of first order
\begin{equation} \label{wkb}
\frac{i Q(r_1)}{\sqrt{2 Q^{(2,*)}}} = n + \frac{1}{2}
\end{equation}
where $Q=\omega^2-V$, $r_1$ minimizes $Q$, and $Q^{(2,*)}$ is the second derivative of $Q$ with respect to the tortoise coordinate evaluated at $r_1$.

In the same regime, $l \gg 1$, our formula becomes approximately
\begin{equation} \label{analytical}
\frac{\omega_{th}}{\kappa} \simeq l \sqrt{\frac{r_c-r_H}{2 \kappa r_H^2}} - i \left(n+\frac{1}{2}\right) 
\end{equation}
Clearly both expressions have the same structure, namely there is a positive real part proportional to $l$ without $n$ dependence, and a negative imaginary part without $l$ dependence, and with precisely the same $n$ dependence, $(n+1/2)$. In addition, we can see that the Lyapunov exponent coincides with the surface gravity, $\kappa = |\lambda_L|$. 

We now explicitly demonstrate that for the five-dimensional
black hole. First we note that the parameters $M,a$ of the black hole are related to the event and cosmological horizons as follows
\begin{equation}
a^2 = r_H^2 + r_c^2
\end{equation}
\begin{equation}
M = \frac{r_H^2 r_c^2}{2 (r_H^2 + r_c^2)}
\end{equation}
and that the solution of the algebraic equation (\ref{algebraic}) in the $D=5$ case is just $r_1=2 \sqrt{M}$. Then all quantities of interest, namely surface gravity $\kappa$, Lyapunov exponent $\lambda_L$ as well as angular velocity $\Omega_c$ are computed to be
\begin{equation}
\kappa = \frac{r_c^2 - r_H^2}{r_H (r_H^2 + r_c^2)}
\end{equation}
\begin{equation}
\Omega_c = \frac{r_c^2 - r_H^2}{2 r_H r_c (r_H^2 + r_c^2)^{1/2}}
\end{equation}
\begin{equation}
|\lambda_L| = \sqrt{2} \Omega_c
\end{equation}
which in the nearly extremal case $a^2 \rightarrow 8M$, $r_c \simeq r_H$, become $\sqrt{2} \Omega_c \simeq \kappa \simeq (r_c-r_H)/r_H^2$. Therefore, both expressions (\ref{eikonal}) and (\ref{analytical}) become 
\begin{equation}
\frac{\omega_{l \gg 1}}{\kappa} \simeq \frac{l}{\sqrt{2}} - i \left(n+\frac{1}{2}\right) 
\end{equation}
and this concludes our derivation.

Finally, in table~\ref{table:Comparison} below we show the real and imaginary part of the QN modes in the eikonal limit for $l=100,120,140,160,180,200$ and $n=0,1,2,3,4,5$. In the parenthesis the WKB results are shown.

\begin{table}[ht!]
\tbl{Comparison between WKB method (in parenthesis) and our formula for $D=5, M=1, a=2.82845$.}
{
\begin{tabular}{l | l | l | l}
		Degree $l$ & $Re(\omega)$ & Overtone $n$ & $Im(\omega)$ \\
		\hline
		\hline
		100 & 0.144 (0.142) & 0 & -0.001 (-0.001) \\
		120 & 0.173 (0.171) & 1 & -0.003 (-0.003) \\
		140 & 0.201 (0.199) & 2 & -0.005 (-0.005) \\
		160 & 0.230 (0.228) & 3 & -0.007 (-0.007) \\
		180 & 0.258 (0.256) & 4 & -0.009 (-0.009) \\
		200 & 0.287 (0.284) & 5 & -0.011 (-0.011) \\
		
\end{tabular}
	\label{table:Comparison}
}
\end{table}
The numerical values obtained show that regarding the imaginary part the agreement is excellent, while regarding the real part the agreement is of the order of $1 \%$.


\section{Conclusions}

We have studied the electromagnetic quasinormal modes of the nearly-extremal Schwarzschild-de Sitter black hole in $D > 4$ dimensions, and we have computed the spectrum analytically. The real part of the frequencies depends on the degree of angular momentum only, while the imaginary part on the overtone number only. Known results from previous works are recovered as special cases. Finally, the WKB semi-analytical approximation was used to verify our results.


\section*{Acknowlegements}

The author is grateful to the anonymous reviewers for useful comments and suggestions. G. P. thanks the Funda\c c\~ao para a Ci\^encia e Tecnologia (FCT), Portugal, for the financial support to the Center for Astrophysics and Gravitation-CENTRA,  Instituto Superior T\'ecnico,  Universidade de Lisboa,  through the Grant No. UID/FIS/00099/2013.



\begin{thebibliography}{99}
\bibitem{wheeler} T.~Regge and J.~A.~Wheeler,
  Phys.\ Rev.\  {\bf 108} (1957) 1063.

\bibitem{zerilli1} F.~J.~Zerilli,
  Phys.\ Rev.\ Lett.\  {\bf 24} (1970) 737.
  
\bibitem{zerilli2} F.~J.~Zerilli,
  Phys.\ Rev.\ D {\bf 2} (1970) 2141.
  
\bibitem{zerilli3} F.~J.~Zerilli,
  Phys.\ Rev.\ D {\bf 9} (1974) 860.

\bibitem{moncrief} V.~Moncrief,
  Phys.\ Rev.\ D {\bf 12} (1975) 1526.

\bibitem{teukolsky} S.~A.~Teukolsky,
  Phys.\ Rev.\ Lett.\  {\bf 29} (1972) 1114.

\bibitem{monograph} S.~Chandrasekhar,
  ``The mathematical theory of black holes,''
  OXFORD, UK: CLARENDON (1985) 646 P.
  
\bibitem{ligo1} B.~P.~Abbott {\it et al.} [LIGO Scientific and Virgo Collaborations],
  Phys.\ Rev.\ Lett.\  {\bf 116} (2016) no.6,  061102
[arXiv:1602.03837 [gr-qc]].

\bibitem{ligo2} B.~P.~Abbott {\it et al.} [LIGO Scientific and Virgo Collaborations],
  Phys.\ Rev.\ Lett.\  {\bf 116} (2016) no.24,  241103
[arXiv:1606.04855 [gr-qc]].

\bibitem{ligo3} B.~P.~Abbott {\it et al.} [LIGO Scientific and VIRGO Collaborations],
  Phys.\ Rev.\ Lett.\  {\bf 118} (2017) no.22,  221101
[arXiv:1706.01812 [gr-qc]].

\bibitem{Yunes} N.~Yunes, K.~Yagi and F.~Pretorius,
  Phys.\ Rev.\ D {\bf 94} (2016) no.8,  084002
  [arXiv:1603.08955 [gr-qc]].

\bibitem{Roman} R.~Konoplya and A.~Zhidenko,
  Phys.\ Lett.\ B {\bf 756} (2016) 350
  [arXiv:1602.04738 [gr-qc]].

\bibitem{review1} K.~D.~Kokkotas and B.~G.~Schmidt,
  Living Rev.\ Rel.\  {\bf 2} (1999) 2
[gr-qc/9909058].

\bibitem{review2} E.~Berti, V.~Cardoso and A.~O.~Starinets,
  Class.\ Quant.\ Grav.\  {\bf 26} (2009) 163001
  [arXiv:0905.2975 [gr-qc]].
  
\bibitem{review3} R.~A.~Konoplya and A.~Zhidenko,
  Rev.\ Mod.\ Phys.\  {\bf 83} (2011) 793
  [arXiv:1102.4014 [gr-qc]].

\bibitem{sn1} A.~G. Riess {\em et~al.}, ``{Observational evidence from supernovae for an
  accelerating universe and a cosmological constant},'' {\em Astron. J.},
  vol.~116, pp.~1009--1038, 1998.

\bibitem{sn2} S.~Perlmutter {\em et~al.}, ``{Measurements of Omega and Lambda from 42 high redshift supernovae},'' {\em Astrophys. J.}, vol.~517, pp.~565--586, 1999.

\bibitem{adscft1} J.~M. Maldacena, ``{The Large N limit of superconformal field theories and supergravity},'' {\em Int. J. Theor. Phys.}, vol.~38, pp.~1113--1133, 1999.
\newblock [Adv. Theor. Math. Phys.2,231(1998)].

\bibitem{adscft2} I.~R. Klebanov, ``{TASI lectures: Introduction to the AdS / CFT
  correspondence},'' in {\em {Strings, branes and gravity. Proceedings,
  Theoretical Advanced Study Institute, TASI'99, Boulder, USA, May 31-June 25,
  1999}}, pp.~615--650, 2000.
  
\bibitem{kaluza} T.~Kaluza,
  Sitzungsber.\ Preuss.\ Akad.\ Wiss.\ Berlin (Math.\ Phys.\ ) {\bf 1921} (1921) 966.
  
\bibitem{klein} O.~Klein,
  Z.\ Phys.\  {\bf 37} (1926) 895
   [Surveys High Energ.\ Phys.\  {\bf 5} (1986) 241].

\bibitem{nilles} H.~P.~Nilles,
  Phys.\ Rept.\  {\bf 110} (1984) 1.
  
\bibitem{ST1} M. B. Green, J. H. Schwarz and E. Witten, \textit{Superstring Theory, Vol. 1 \& 2}, Cambridge University Press, Cambridge, England, 2012.

\bibitem{ST2} J. Polchinski, \textit{String Theory, Vol. 1 \& 2}, Cambridge University Press, Cambridge, England, 2005.

\bibitem{basic} V.~Cardoso and J.~P.~S.~Lemos,
  Phys.\ Rev.\ D {\bf 67} (2003) 084020
  [gr-qc/0301078].
  
\bibitem{paper2} A.~Zhidenko,
  Class.\ Quant.\ Grav.\  {\bf 21} (2004) 273
  [gr-qc/0307012].

\bibitem{paper3} R.~A.~Konoplya and A.~Zhidenko,
  JHEP {\bf 0406} (2004) 037
  [hep-th/0402080].
  
\bibitem{paper4} R.~A.~Konoplya,
  Phys.\ Rev.\ D {\bf 66} (2002) 044009
  [hep-th/0205142].

\bibitem{paper5} V.~Cardoso, R.~Konoplya and J.~P.~S.~Lemos,
  Phys.\ Rev.\ D {\bf 68} (2003) 044024
  [gr-qc/0305037].

\bibitem{wkb1} S.~Iyer and C.~M.~Will,
  Phys.\ Rev.\ D {\bf 35} (1987) 3621.

\bibitem{wkb2} R.~A.~Konoplya,
  Phys.\ Rev.\ D {\bf 68} (2003) 024018
[gr-qc/0303052].

\bibitem{KZ} R.~A.~Konoplya and A.~Zhidenko,
  Nucl.\ Phys.\ B {\bf 777} (2007) 182
  [hep-th/0703231 [HEP-TH]].

\bibitem{molina} C.~Molina,
  Phys.\ Rev.\ D {\bf 68} (2003) 064007
  [gr-qc/0304053].
  
\bibitem{master} H.~Kodama and A.~Ishibashi,
  Prog.\ Theor.\ Phys.\  {\bf 110} (2003) 701
  [hep-th/0305147].

\bibitem{ortega1} A.~Lopez-Ortega,
  Gen.\ Rel.\ Grav.\  {\bf 38} (2006) 1747
  [gr-qc/0605034].
  
\bibitem{ortega2} A.~Lopez-Ortega,
  Gen.\ Rel.\ Grav.\  {\bf 40} (2008) 1379
  [arXiv:0706.2933 [gr-qc]].
  
\bibitem{ortega3} A.~Lopez-Ortega,
  Gen.\ Rel.\ Grav.\  {\bf 38} (2006) 1565
  [gr-qc/0605027].
  
\bibitem{crispino} L.~C.~B.~Crispino, A.~Higuchi and G.~E.~A.~Matsas,
  Phys.\ Rev.\ D {\bf 63} (2001) 124008
   Erratum: [Phys.\ Rev.\ D {\bf 80} (2009) 029906]
  [gr-qc/0011070].

\bibitem{potential} G.~Poschl and E.~Teller,
  Z.\ Phys.\  {\bf 83} (1933) 143.

\bibitem{ferrari} V.~Ferrari and B.~Mashhoon,
  Phys.\ Rev.\ D {\bf 30} (1984) 295.
  
\bibitem{eikonal} V.~Cardoso, A.~S.~Miranda, E.~Berti, H.~Witek and V.~T.~Zanchin,
  Phys.\ Rev.\ D {\bf 79} (2009) 064016
  [arXiv:0812.1806 [hep-th]].
\end{thebibliography}
\end{document}